# Design parameters of free-form color routers for subwavelength pixelated CMOS image sensors

Sanmun Kim[1†], Chanhyung Park[1†], Shinho Kim[1], Haejun Chung[2], and Min Seok Jang[1]*

[1] *School of Electrical Engineering, Korea Advanced Institute of Science and Technology, Daejeon 34141, Republic of Korea*
[2] *School of Electrical Engineering, Hanyang University, Seoul 04763, Republic of Korea*
*\* jang.minseok@kaist.ac.kr*

## ABSTRACT

Metasurface-based color routers are emerging as next-generation optical components for image sensors, replacing classical color filters and microlens arrays. In this work, we report how the design parameters such as the device dimensions and refractive indices of the dielectrics affect the optical efficiency of the color routers. Also, we report how the design grid resolution parameters affect the optical efficiency and discover that the fabrication of a color router is possible even in legacy fabrication facilities with low structure resolutions.

## Introduction

The major optical components in classical complementary metal–oxide–semiconductor (CMOS) image sensors are a microlens array and color filter. A microlens focuses the incident light on the photodiode and the color filter blocks the light of unwanted wavelength (Figure 1a). However, such geometric-optics-based configuration is limited to image sensors with pixel sizes relatively large compared to the wavelength [1,2]. Recent developments in CMOS image sensors brought the subpixel size down to 0.56 μm [3], reaching down to the boundary of the geometric optics and wave optics. Further miniaturization will likely cause a failure in design approaches based on geometric optics. Furthermore, a decrease in the subpixel size has led to a reduction of light energy per subpixel leading to poor image quality. Metasurface-based color routers are being investigated as a candidate for substituting microlenses and color filters due to their high optical efficiency. Instead of filtering out lights of unwanted wavelength, the color router guides the incoming light to the corresponding subpixels, thus opening the way to utilize light incident on the entire image sensor area. Compared to the conventional CMOS image sensors whose subpixels only utilize either quarter (red and blue) or half (green) of the incident light, the color routers can in principle exhibit 2 to 4 times higher optical efficiencies.

Metasurface-based color router is configured by allocating dielectrics of different refractive indices inside the design region. This is a typical freeform optimization problem involving high degrees of freedom (DoF). There have been many attempts to solve such optimization problems. The first measurement data in the visible range was reported by Miyata et al. [4]. The authors used a conventional library-based meta-atom method to design a single-layer color router. Although a library-based method has a significantly constrained design space, the fabricated device already showed superior performance compared to the classical color filter. Similar work was followed by Zou et al. who designed a single-layer freeform color router using the genetic algorithm and measured its performance in the visible range [5]. Investigation on multilayer devices has also been reported. Although a multilayer device tends to show higher performance, the design optimization of a multilayer device is much harder than that of a single-layer device due to the large DoF. A typical approach to handle this large DoF is to utilize local figure-of-merit gradients on design variables obtained through auto-gradient calculation [6] or the adjoint method [7]. Zhao et al. [8] and Catrysse et al. [9] optimized high-complexity color routers in 2D and 3D space using auto-gradient calculations. The design space is meshed with ultra-fine grids, and the authors were able to obtain a device design with near-perfect efficiency. Another pioneering work was done by Camayd-Muñoz et al. where an adjoint-based method was applied to design a 3D device with a higher fabricability . Despite the rapidly growing field [10-16], there has not been any systematic investigation into the choice of device design parameters. The choice of design parameters, such as the device height or selection of refractive index, has a critical effect on the final optimized devices. Until now, the choice of such parameters was based on simple deductions such as Fabry-Perot resonance conditions or even worse, based on the computational resource availability [15].

In this work, we outline the effect of design parameters on the optimized optical efficiency of a color router. The result shows that there exist optimal ranges of both structural parameters and the optical index contrast of constituting materials, and more interestingly, their optimal ranges are correlated with each other. We also investigate how the spatial grid size and the number of grid layers affect the optical efficiency and demonstrate that a sufficiently high-performing device can be obtained even with a large cell size if a sufficient number of layers are deposited. This highlights the important role that the choice of design parameters plays in determining the device's performance.

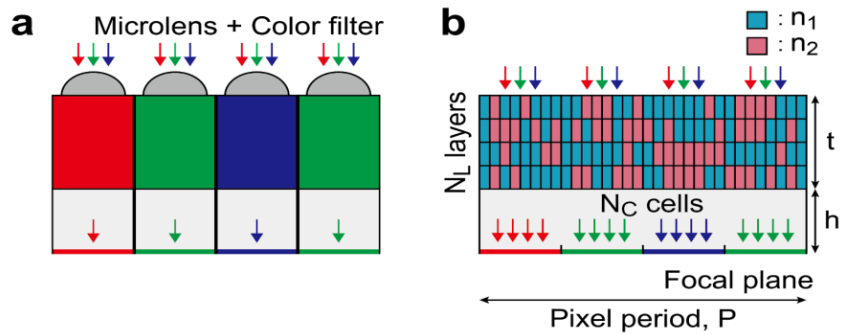

**Figure 1.** (a) A simplified diagram of a conventional CMOS image sensor consisting of a microlens array and color filter. (b) A schematic diagram of a color router. The design area ($P \times t$) is gridded into a grid of $N_C \times N_L$, and refractive indices $n_1$ and $n_2$ are allocated to each cell for color routing. Four arrows at the focal plane of the color router imply that an ideal color router can have a four-fold increase in optical efficiency compared to the conventional design.

## Results

As schematically shown in Figure 1b, a color router deflects the incident light to its corresponding subpixel area. Instead of forming a lens-like structure, the design area is gridded into rectangular cells, and each cell is filled with a selection of two different dielectrics. The design parameters for the 2D color routers can be classified into two categories: physical parameters and spatial resolution parameters. The physical parameters include color router period ($P$), thickness ($t$), the position of the focal plane ($h$), and the refractive indices of the two composing dielectrics ($n_1$ and $n_2$). The spatial resolution, determined by the number of grid layer $N_L$ and the grid elements in a layer $N_C$, defines how the design area is gridded into cells of equal shape. Consequently, the design problem possesses $N_L \times N_C$ degrees of freedom and thus the number of possible structures is $2^{N_L N_C}$. The default values of each design parameter are given in Table 1. As the transition from geometric optics to wave optics occurs for geometries with characteristic lengths comparable to or smaller than the wavelength, the color router configured with the default design parameters lies within the wave optics regime.

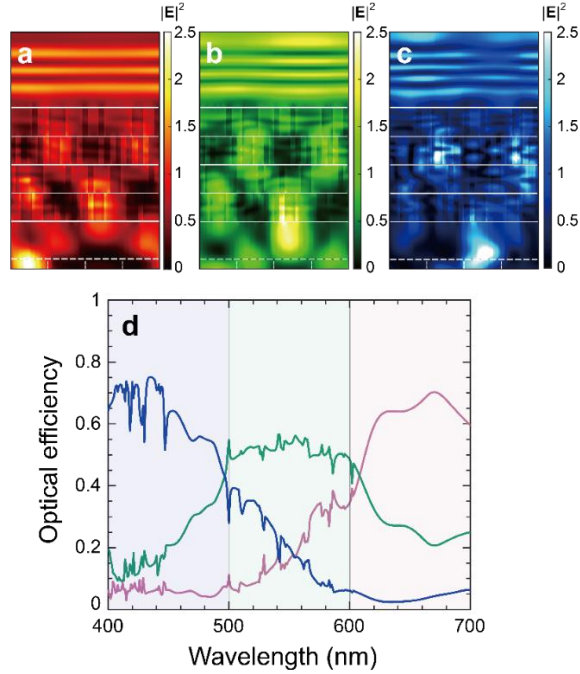

**Figure 2.** The electric field intensity profile inside the optimized device is given in Figure 1b for a normally incident light of (a) λ = 650 nm, (b) λ = 550 nm, (c) λ = 450 nm. The depicted field distribution is the average of transverse electric and transverse magnetic polarized light. (d) Optical efficiency spectra of the same device. The default design parameters in Table 1 are used. The average optical efficiencies between 400 nm to 700 nm are 58.29%.

In this work, we define the optical efficiency $\eta(\lambda)$ using the electric field intensity at the focal plane (denoted by the dashed line in Figure 2a-c).

$$\eta_{R,G,B}(\lambda) = \frac{1}{2} \sum_{i=\text{TE,TM}} \frac{\int_{x_1}^{x_2} |\mathbf{E}(\lambda,i)|^2 dx}{\int_0^P |\mathbf{E}(\lambda,i)|^2 dx} \times T(\lambda,i)$$

Here, $E$ is the electric field at the focal plane, and $T$ is the transmittance. Electric field distribution and the total transmittance are calculated with RETICOLO, a rigorous coupled-wave analysis package [17]. $x \in (x_1, x_2)$ defines the area of the subpixel of interest. For simplicity, we assume that the wavelength range required for red (R), green (G), and blue (B) subpixels are 600 nm - 700nm, 500 - 600 nm, and 400 nm - 500 nm, respectively. Throughout the work, a normally incident light is assumed, and the optical efficiency is averaged between both transverse electric (TE) and transverse magnetic (TM) polarizations. Figure 2(a-c) shows the electric field intensity distribution inside an optimized color router with the default design parameters listed in Table 1. The optical efficiency $\eta_{R,G,B}(\lambda)$ of the same device is shown as red, green, and blue curves in Figure 2d. Both the field distributions and the optical efficiency plot clearly show that the intensity of light is concentrated at the corresponding subpixel area on the focal plane.

In a conventional Bayer-type image sensor, a pixel consists of two green subpixels and one subpixel for red and blue, respectively. In order to account for such a subpixel ratio, we include two green subpixels in one period of a 1D image sensor. The default arrangement of the subpixels in this paper was set to RGBG as the design is periodic and the wavelength of the green light is in between red and blue (Figure 1b). In Supplementary Figure S1, we compare the optical efficiency between the RGBG subpixel arrangement and the RGGB subpixel arrangement. As Supplementary Figure S1 suggests, the arrangement of subpixels has a marginal effect on the device performance in terms of optical efficiency and crosstalk.

**Table 1. The default design parameters used in this work**

| Design parameter | Value |
| --- | --- |
| Pixel period, $P$ | 1 μm (equivalent to subpixel size of 0.25 μm) |
| Thickness of the color router, $t$ | 1.5 μm |
| Refractive index of dielectric 1, $n_1$ | 1.5 |
| Refractive index of dielectric 2, $n_2$ | 2.0 |
| Position of the focal plane, $h$ | 0.5 μm |
| Number of grid layers ($N_L$) | 4 |
| Number of cells in a layer ($N_C$) | 32 |

To understand how the design parameters of a color router affect its performance, we optimize the device geometry for various choices of design parameters. For given device design parameters, a conventional genetic algorithm with elitism is performed to obtain the optimal dielectric distribution in the grids. The optimization is configured with a population size of 200, and 100 epochs. The genotype of the individuals in the gene pool is represented by a binary array with array dimensions equal to $N_C$ and $N_L$. The goal of the optimization is to maximize the average optical efficiency, $\bar{\eta} = (\bar{\eta}_R + \bar{\eta}_G + \bar{\eta}_B)/3$, where $\bar{\eta}_{R,G,B}$ are the wavelength-averaged optical efficiencies obtained by averaging $\eta_{R,G,B}(\lambda)$ over the wavelength range corresponding to the subpixel type. During the optimization process, the optical efficiencies were averaged over thirty wavelength points (405 nm, 415 nm, … 695 nm) to reduce the computational cost, but the reported $\bar{\eta}$ were averaged over with much finer wavelength points (400 nm, 401 nm, … 700 nm). As shown in Figure S2 in the Supplementary Information, the difference between 30 and 301 wavelength-point averaging is not significant (around 0.02 for $(N_L, N_C) = (8, 64)$ and 0.01 for $(N_L, N_C) = (4, 32)$).

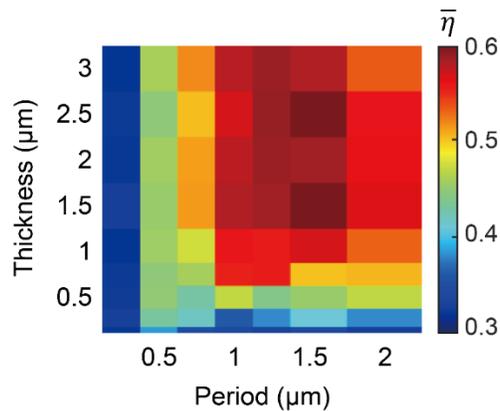

**Figure 3.** Effect of device period ($P$) and color router thickness ($t$) on the optical efficiency of a color router. $N_L$ and $N_C$ are fixed to (4, 32). The design parameters stated in Table 1 are used except for $P$, and $t$.

The advantage of substituting microlens and color filters with metasurface-based color routers becomes clear for sub-micron image sensors. Hence, we first investigate the effect of the physical dimensions of devices on the optical efficiency of the router. It should be noted that the subpixel size of the 2D color router is a quarter of the device period, $P$. In comparison to a Bayer-type image sensor array, a 2D color router extends infinitely in the $y$-direction so the pixel size is defined as the width of each subpixel in the $x$-direction. The pixel size of the color router with the default design parameter is 0.25 μm, which is less than half the size of the smallest commercially available image sensor of ~0.56 μm [3]. Figure 3 shows how the optimized $\bar{\eta}$ varies depending on the period $P$ and the thickness $t$ while all the other design parameters including DoF and refractive indices are fixed to their default values. For the devices with a deep subwavelength period of $P = 0.25$ μm, the optimized average optical efficiencies are around the trivial value of 33%, which can be achieved with a simple antireflection layer. When $P \geq 0.5$ μm, the color routers start to show meaningful performance. At a given $P$, the device performance monotonically increases and saturates as the thickness $t$ increases. The saturation point of $t$ for 0.75 μm $\leq P \leq$ 2 μm is around 1.5 μm, and thus we set $t = 1.5$ μm as the default value. We note that, however, the saturation point of $t$ can vary as a function of the other design parameters. At a fixed $t$, the optimized $\bar{\eta}$ does not monotonically increase with $P$ but has a specific optimal value. This result is reasonable since it becomes increasingly difficult to route incident light over a longer lateral distance within a given thickness.

The position of the focal plane from the color router, $h$, is a similar physical design parameter to $P$ and $t$, which also defines the physical dimension of the device. The dependence of the focal plane position on the optical efficiency is shown in Supplementary Figure 3. In a periodic grating, the modes with a high lateral wavenumber cannot be extracted in the far field. Hence, as the focal plane of the color router is located further from the meshed region, the device is expected to have a lower efficiency due to the loss of near field. The sharp drop in optical efficiency for $h >$ 1 μm in Supplementary Figure S3 agrees with this expectation.

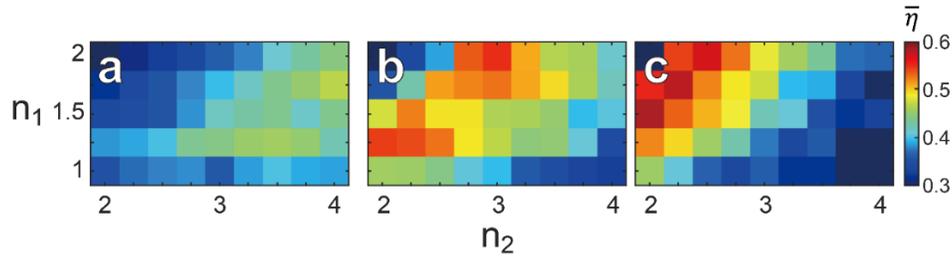

**Figure 4.** Effect of refractive indices on the optical efficiency for different device thicknesses, $t$. Optimization based on a genetic algorithm was carried out for color routers with thickness (a) $t = 0.1$ μm, (b) $t = 0.5$ μm, and (c) $t = 1.5$ μm. In each color plot, the lower refractive index $n_1$ is changed from 1 to 2 with a step size of 0.25, and $n_2$ is swept from 2 to 4 with the same step size, 0.25. Each square represents the optimized efficiency obtained with the genetic algorithm. The maximum efficiency in each case is (a) 46.86%, (b) 54.98%, (c) 58.25%. Except for $t$, $n_1$, and $n_2$, the design parameters given in Table 1 are used.

The refractive indices of the composing dielectric materials are another critical factor determining optimal efficiency. In previous works, the selection of a color router was based on simple relations such as the Fabry-Perot resonance condition [15]. Those relations only provide order-of-magnitude estimates. In this work, we tune the design parameters ($t$, $n_1$, $n_2$) to find the global trend in optimized optical efficiency. For the sake of simplicity, we assume that the dielectrics filling each grid are dispersionless and have refractive indices of $n_1$ and $n_2$, where $n_1 \leq n_2$ is assumed throughout the work. The default values of ($n_1$, $n_2$) are (1.5, 2.0), which is similar to the refractive indices of silica and silicon nitride. Our analyses reveal that, unlike other nanophotonic devices such as metalens whose device performance monotonically increases with the refractive index contrast [18-20], color routers have a distinct relation between the optimal refractive index contrast and the thickness of the device. When all the other parameters are fixed to their default values, the optimal index contrast values, $n_1 - n_2$, are found to be 2.25, 1, and 0.5 for t = 0.1, 0.5, and 1.5 μm, respectively, as illustrated in Figure 4. We speculate that trend could be attributed to the fact that the maximum achievable vertical optical path length difference is determined by the product of optical index contrast and the thickness of the device.

The choice of DoF is important in both computational and experimental aspects. On the computational side, the design space grows exponentially with the DoF, and the computational load required for optimization grows accordingly. Popular approaches for tackling high DoF problems are through the adjoint gradient, which provides the gradient of FoM with respect to change in the refractive index of every element in the design space [21-29], or through machine learning methods [30-38]. In our work, we limit the DoF to the order of hundreds so that the optimization problem is solvable using the classical genetic algorithm [39-43]. On the other hand, the DoF is directly related to the fabrication feasibility of the device. The number of layers, $N_L$, determines the number of deposition steps, and the number of cells in a layer, $N_C$, affects the minimum feature size. Despite its importance, previous works on metasurface based color routers mostly lack investigations on DoF. In this work, we fix the values of the other design parameters including the device thickness, and change $N_L$ and $N_C$ to isolate the effect arising from the device dimension change. $N_L$ and $N_C$ are chosen to be integer powers of 2. This implies the existence of trivial monotonicity. For example, a set of every possible combination with ($N_L = 1$, $N_C = 8$) is a subset of ($N_L = 4$, $N_C = 16$) so the optical efficiency of the latter must be equal to or greater than the previous one if the optimization converges to the global optimum. Since the number of possible combinations is sufficiently low for device designs with DoF ≤ 16, an exhaustive search was carried out for the corresponding conditions. For device designs with DoF ≥ 32, the previously-described genetic algorithm was carried out.

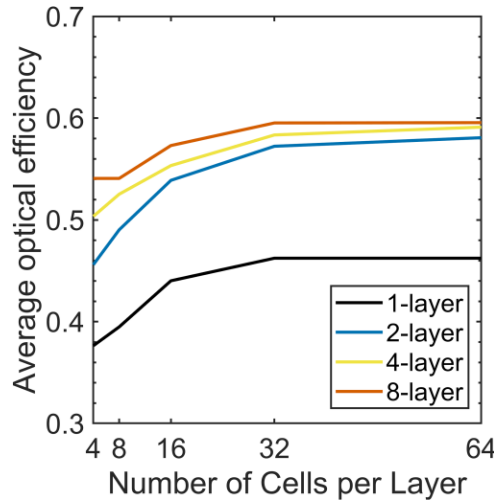

**Figure 5.** Optimized optical efficiency calculated for multiple DoF configurations. The optical efficiency saturates to ~60%. Except for $N_L$ and $N_C$, the design parameters given in Table 1 are used.

Figure 5 shows the optimized results for each $N_L$ and $N_C$ pair. In the figure, the trivial monotonic relation in the optimized efficiency is observed. Regardless of the number of layers, the optimal $\bar{\eta}$ almost saturates when $N_C \geq 32$, which corresponds to the minimum feature size of ~31 nm. The optimal $\bar{\eta}$ asymptotically approaches ~ 60% for the default physical parameters. It is important to note that, the number of layers $N_L$ plays a pivotal role in determining the device performance. For example, even with $N_C = 4$ (minimum feature size of 250 nm), it is possible to achieve the average optical efficiency of ~ 54% (about 90% of the maximum achievable efficiency) by having 8 layers. The designs of color routers for different DoF conditions are displayed in Supplementary Section 4. For low-efficiency devices, a line of reflection symmetry exists at the center of the red and blue subpixel. This line of reflection symmetry originates from RGBG subpixel arrangement which is also symmetric with respect to that line. However, such reflection symmetry isn't observed in the optimized devices. Lack of symmetry in the optimized devices implies that the enforcement of trivial symmetry conditions on the device design does not always lead to better performance.

## Discussion

In conclusion, we systematically analyze the dependence of color router performance on various design parameters by leveraging numerical device optimization methods based on a genetic algorithm. We discover that the average optical efficiency of a color router with a micron-scale form factor can be up to ~60%, whereas the classical microlens and color filter configuration can have optical efficiency up to 25% for red and blue and 50% for green. We show that it is not always beneficial to have a larger pixel if the thickness of the device is limited and there exist optimal refractive index pairs for composing dielectrics for a given device thickness. Unlike the case of metalens, the optical efficiency drops when the refractive index contrast becomes greater than the optimal value. We also report that the device performance can be greatly increased while maintaining a relatively large feature size by having multiple layers in the design scheme. We anticipate that the qualitative trend seen in the 2D color router design parameter tuning would be repeated for Bayer-type 3D color routers, although the optimal values may differ due to the introduction of the additional dimension. Our results will serve as a design guideline for the future development of free-form metasurface-based color routers for deep sub-micron image sensors.

## Author contributions

Sanmun K. and M.S.J conceived the ideas. Sanmun K., C. P., and Shinho K. developed the optical simulation model and the optimization algorithms. Sanmun K. and M.S.J. conducted a detailed analysis of the optimization results. M.S.J. supervised the project. The manuscript was mainly written by Sanmun K., H. C., and M.S.J. with the contributions of all authors.

## Data and Code availability

The optimization code can be accessed openly on Sanmun Kim's GitHub (https://github.com/chocopi2718/colorRouter2D).

## Acknowledgements

This research was supported by the MOTIE (Ministry of Trade, Industry & Energy 1415180303 and KSRC (Korea Semiconductor Research Consortium) 20019357 support program for the development of the future semiconductor device.

## Competing Interests

The authors declare no conflicts of interest.

## Supplemental Document

The following files are available free of charge.

Supplementary information (Supplementary information.pdf)

S1. Arrangement of subpixels: RGBG vs RGGB

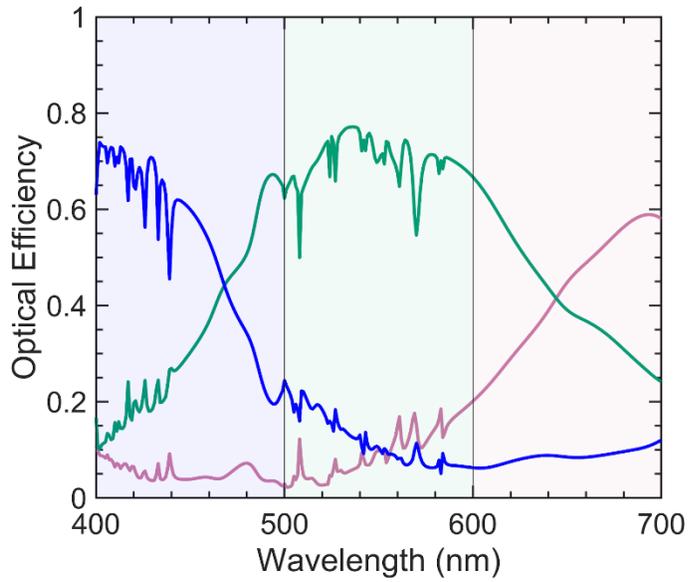

Figure S1. [The optical efficiency spectrum of RGGB configuration] The average optical efficiency and optical crosstalk of this device are 54.95% and 18.69%, respectively. The optical efficiency and crosstalk of the device in Figure 2d with RGBG arrangement are 58.29% and 16.84%. The arrangement of subpixels only has a limited effect on optical efficiency.

S2. Number of sampling wavelengths vs accuracy

The optical efficiencies reported in this work are calculated by averaging optical efficiencies over 301-wavelength points (400 nm, 401 nm, …, 700 nm). Since we are utilizing a rigorous coupled-wave analysis solver, the computational cost is proportional to the number of wavelength points. Hence, during the optimization process, the optical efficiency is sampled from thirty wavelength points (405 nm, 415 nm, …, 695 nm). In this section, we report how the error in average optical efficiency changes as a function of the number of sampling points. The error in optical efficiency is defined as the discrepancy between the 301-wavelength points averaged value and the N-wavelength points averaged value.

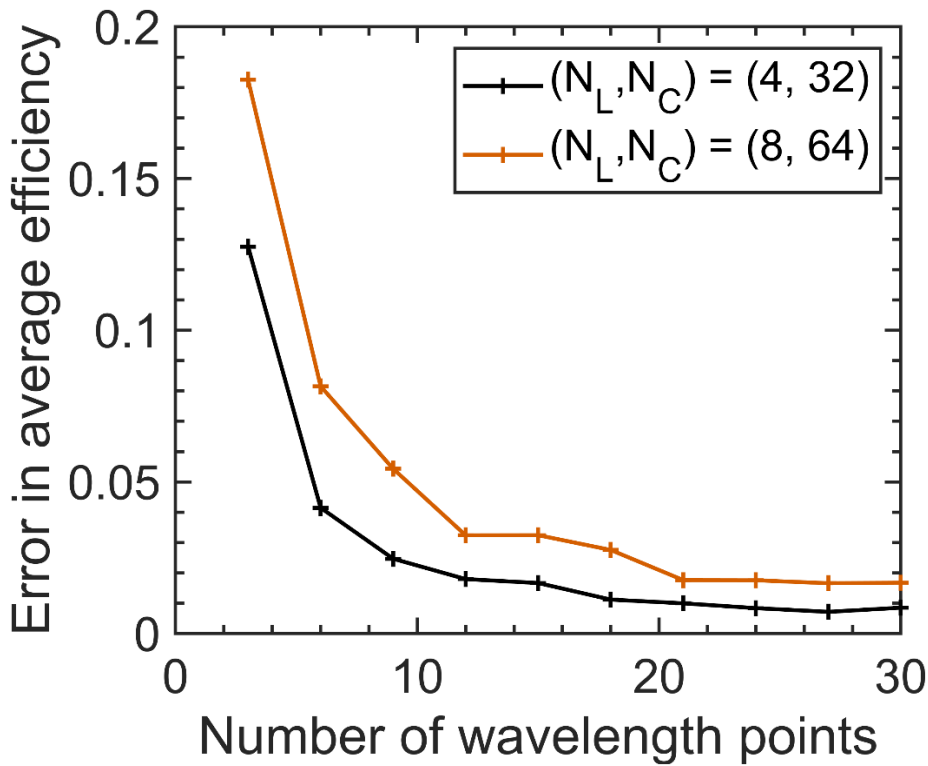

Figure S2. [The error in average efficiency as a function of sample point number] Data in the plot is the root-mean-squared value of $|\overline{\eta_{301}} - \overline{\eta_{30}}|$. Devices with a higher DoF require more sampling points as peaks and dips are likely to be sharper for a more complex device.

S3. Position of the focal plane vs optical efficiency.

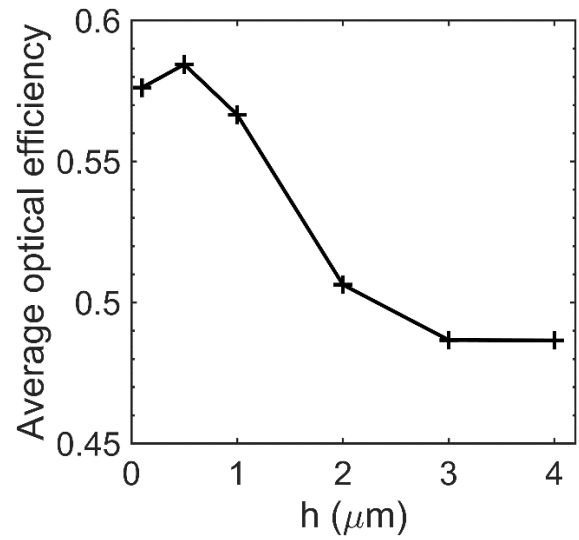

Figure S3. [Optical efficiency plot as a function of focal plane position (h)] The averaged optical efficiency drops with h due to the loss of near-field contribution.

S4. Exemplary device designs for various DoF.

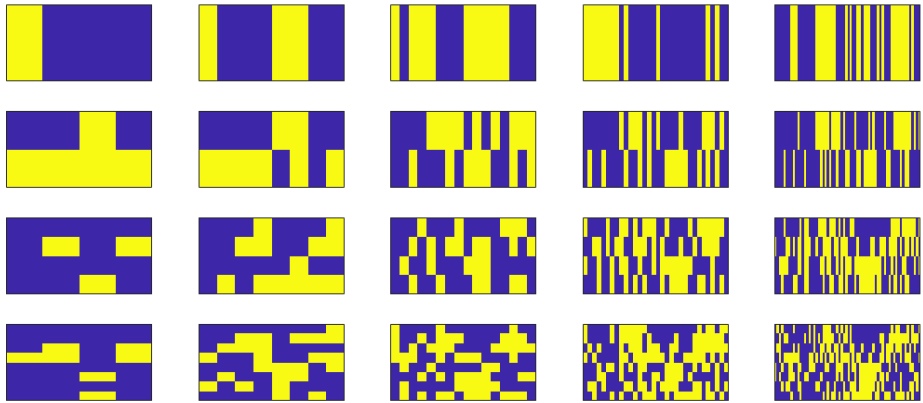

Figure S4. [Exemplary device designs for various DoF] Region colored with yellow and violet each represents space filled with $n_2$ and $n_1$, respectively.